# Crossover between liquid-like and gas-like behaviour in CH$_4$ at 400 K

D. Smith[1,2], M.A. Hakeem[1], P. Parisiades[3,4], H.E. Maynard-Casely[5], D. Foster[1], D. Eden[1], D.J. Bull[1], A.R.L. Marshall[2], A.M. Adawi[2], R. Howie[6,7], A. Sapelkin[8], V.V. Brazhkin[9] and J.E. Proctor*[1,2,10]

1 Materials and Physics Research Group, School of Computing, Science and Engineering, University of Salford, Manchester M5 4WT, United Kingdom

2 School of Mathematics and Physical Sciences, University of Hull, Hull HU6 7RX, United Kingdom

3 European Synchrotron Radiation Facility, Beamline ID27, BP 220, Grenoble, France

4 IMPMC, Université Pierre et Marie Curie, 4 place Jussieu, 75005 Paris, France

5 Australian Nuclear Science and Technology Organisation, Locked Bag 2001, Kirrawee DC, NSW, 2232, Australia

6 SUPA, School of Physics and Centre for Science at Extreme Conditions, University of Edinburgh, Edinburgh EH9 3JZ, United Kingdom

7 Center for High Pressure Science & Technology Advanced Research (HPSTAR), Shanghai 201203, P.R. China

8 School of Physics and Astronomy, Queen Mary University of London, London E1 4NS, United Kingdom

9 Institute for High Pressure Physics, RAS, 108440 Troitsk, Moscow, Russia

10 Photon Science Institute and School of Electrical and Electronic Engineering, The University of Manchester, Oxford Road, Manchester M13 9PL, United Kingdom

* Corresponding author j.e.proctor@salford.ac.uk

**We report experimental evidence for a crossover between a liquid-like state and a gas-like state in fluid methane ($CH_4$). This crossover is observed in all of our experiments, up to 397 K temperature; 2.1 times the critical temperature of methane. The crossover has been characterized with both Raman spectroscopy and X-ray diffraction in a number of separate experiments, and confirmed to be reversible. We associate this crossover with the Frenkel line - a recently hypothesized crossover in dynamic properties of fluids extending to arbitrarily high pressure and temperature, dividing the phase diagram into separate regions where the fluid possesses liquid-like and gas-like properties. On the liquid-like side the Raman-active vibration increases in frequency linearly as pressure is increased, as expected due to the repulsive interaction between adjacent molecules. On the gas-like side this competes with the attractive Van der Waal's potential leading the vibration frequency to decrease as pressure is increased.**

## I. INTRODUCTION

For centuries, the existence of a first-order phase transition, involving a large discontinuous change in density, between the liquid and gas states of matter, has been understood. Applying high pressure shifts this transition to higher temperatures. In 1822 the critical point was discovered[1], a specific pressure ($P_c$) and temperature *($T_c$)* where the first-order phase transition line between liquid and gas states ends. Beyond this point, a discontinuous change in a physical observable distinguishing between a liquid-like and gas-like state in the sample cannot be discerned. Matter in this *P-T* region is therefore described as a "supercritical fluid"; neither a liquid nor a gas.

Theoretical description of the liquid and supercritical fluid phases has been a challenge ever since. Previously the common approach was to treat liquids and supercritical fluids as dense non-ideal gases[2], justified on the basis that liquids and supercritical fluids share important properties with gases; for instance a lack of long-range order. A path on a *P-T* phase diagram can commence in the gas phase, pass above the critical point and end in the liquid phase without encountering a first order phase transition. Whilst no discontinuous change in a physical observable can be observed beyond the critical point, the understanding of the supercritical fluid phase as a dense non-ideal gas has led to the prediction, and observation, of some other crossovers extending a finite distance from the critical point.

There are firstly a group of crossover lines emanating from the exact critical point and extending to higher temperatures and pressures called the Widom lines. The Widom lines are maxima in certain thermophysical properties such as the speed of sound, isochoric and isobaric heat capacities. Further away from the critical point the maxima become more smeared out and the different Widom lines diverge from each other[3][4][5]. No Widom lines are expected to exist beyond $T / T_c \approx$ 2.5 and $P / P_c \approx 3$ [2].

Also beginning in the vicinity of the critical point (but not necessarily emanating from the exact critical point) are the Joule-Thomson inversion curve and Boyle Curve. At temperatures up to ca. $3T_c$ lines of constant enthalpy can be traced across the P-T phase diagram. Each line, as pressure is varied, reaches a maximum temperature. These maxima are linked by the Joule-Thomson inversion curve[7][8]. In a similar manner, the Boyle curve and (at higher pressure) the Amagat curve can be traced out. The nature of these lines is highly relevant to the industrial applications of supercritical fluids, for instance the use of the Joule-Thomson effect in refrigeration technology.

However, as pressure is increased the density of a liquid (or supercritical fluid) becomes close to that of a solid, orders of magnitude larger than the density of a gas. The latent heat of vaporization is usually an order of magnitude larger than the latent heat of fusion, and liquids (unlike gases) can exhibit orientational order similarly to solids. Therefore, instead of treating liquids as dense non-ideal gases, the opposite theoretical approach has also been utilized – treating them similarly to solids, assuming (relatively) closely packed atoms[9][10][11]. This approach was first put forward by Frenkel in several publications leading to *Kinetic Theory of Liquids*[9] and also by Lennard-Jones and

Devonshire[12][13]. In this approach, the molecules (or atoms in an atomic liquid) have definite positions, with occasional vacancies[9][14], which can be treated in a similar manner to vacancies in solids. This is necessary to understand the heat capacities of liquids[11] and to understand why, unlike gases, liquids do support high-frequency shear waves[15].

A key element of Frenkel's theory is the introduction of the liquid relaxation time $\tau_r$. Frenkel proposed that, on short timescales, the molecules in a liquid exhibit only vibrational motion around an equilibrium position, similar to that in a solid. However, the molecules occasionally jump to a new equilibrium position by swapping places with an adjacent molecule or vacancy. The liquid relaxation time $\tau_r$ is the average time that a molecule spends in a certain equilibrium position, between consecutive jumps. On time scales shorter than this, Frenkel proposed, the liquid could support shear waves just like a solid. This ability to support shear waves, with period shorter than the liquid relaxation time $\tau_r$, has been verified experimentally[15] and has allowed accurate prediction of the heat capacities of liquids[11].

Shear waves can therefore be supported in liquids with periods ranging from a minimum of the Debye vibrational period $\tau_0$ up to a maximum of $\tau_r$. The Debye period $\tau_0$ does not change significantly with temperature but $\tau_r$ is expected to decrease by orders of magnitude as temperature increases due to the greater amount of thermal energy available[6]. Therefore, under and isobaric temperature increase commencing in the liquid region (at higher pressure and lower temperature than the critical point), a point is eventually reached where due to the decrease in $\tau_r$ no shear wave can be supported since $\tau_r < \tau_0$. Thus there is a crossover here from liquid-like to gas-like behaviour, which has been named the "Frenkel line"[6].

The Frenkel line is a different phenomenon to the Widom lines[3]. We predict (for instance in $CO_2$, $H_2O$, $CH_4$[16] and Ne[17]) that the Frenkel line at $T_c$ lies at substantially higher pressure than $P_c$, where the Widom lines begin. In $CH_4$ $P_c$ = 4.5 MPa, whilst the Frenkel line at $T_c$ is predicted to occur at 54 MPa[16]. This is because, as Frenkel pointed out[9], the fluid density close to the critical point is far too low for the solid-like description of the liquid as a relatively close packed arrangement of molecules to be valid. As temperature is increased, both the Widom lines and the Frenkel line are crossed at higher pressure but roughly the same density. Thus, at a given temperature, the Frenkel line is always crossed at significantly higher pressure and density than the Widom lines. Furthermore, the Widom lines extend a finite distance from the critical point[3][4] whilst the Frenkel line is expected to extend to arbitrarily high *P-T*.

The liquid-like or gas-like environment affects the vibrations of individual molecules. Many vibrations are measurable using Raman or infra-red spectroscopy and this gives insight into the environment as vibrational frequencies depend on both the density and dynamics of the sample. In a liquid as described by Frenkel, the frequency of vibrations detectable using Raman and infra-red spectroscopy should increase in a linear or slightly sub-linear manner, upon pressure increase, similarly to a solid. This is understood by considering a solid described by the simplest equation of state (EOS), the Murnaghan equation[18][19]:

$$\frac{V(P)}{V_0} = \left[1 + \frac{K_0' P}{K_0}\right]^{-\frac{1}{K_0'}} \quad (1)$$

Here, $K_0$ is the bulk modulus and $K_0'$ its pressure derivative. Provided the application of pressure does not substantially modify the nature of the inter-atomic bonds, one can also relate the sample volume to vibrational frequency $\omega(P)$ using a positive constant Grüneisen parameter $\gamma$ for that mode[20]:

$$\frac{\omega(P)}{\omega_0} = \left[\frac{V(P)}{V_0}\right]^{-\gamma} \quad (2)$$

We can hence predict the variation of $\omega(P)$ with pressure:

$$\frac{\omega(P)}{\omega_0} = \left[1 + \frac{K_0' P}{K_0}\right]^{\frac{\gamma}{K_0'}} \quad (3)$$

Generally, $K_0' \approx 4$ (It is frequently fixed at 4.0[19]), and for typical Raman-active vibrational modes $\gamma \approx 2-4$. Hence $\omega(P)$ increases in a linear or slightly sub-linear manner upon pressure increase. This is almost always observed in dense liquids, amorphous and crystalline solids due to the interaction between adjacent atoms being entirely repulsive under pressure (equation (2) is a consequence of this).

Conversely, in a gas - and in the *P-T* region that we will propose here as the gas-like side of the Frenkel line - there is competition between the repulsive potential and attractive van der Waals forces between adjacent molecules undergoing ballistic motion.  Attractive van der Waals forces dominate in a gas at the lowest pressures and lead the vibrational frequencies to decrease upon pressure increase[21].  Thus, the variation of vibrational frequencies as a function of pressure in an isothermal experiment can display a minimum as one proceeds from the gas-like side to the liquid-like side across the Frenkel line, and be described by a quadratic or cubic function rather than a linear or slightly sublinear function[22][23].

Experimentally, a number of spectroscopy studies have been conducted on gases and supercritical fluids at very low pressure up to *ca.* 500 bars [21][24][25][26], observing the decrease in vibrational frequencies upon pressure increase.  In a few cases these studies have reached a pressure where there is a minimum in the vibrational frequency as the repulsive inter-atomic forces begin to take effect[24] but none have reached the regime on what we now propose as the liquid-like side of the Frenkel line, where repulsion completely dominates over the van der Waals forces.

Additionally, many spectroscopy studies have focussed on solids and dense liquids in the diamond anvil cell (DAC) to characterize the nearly linear increase in vibrational frequencies as a function of pressure in these P-T regions[27][28][29].  However, the low-pressure region where the repulsive part of the inter-atomic potential competes with attractive van der Waal's forces has not been characterized in such studies.  This is partially because it is challenging to control and measure pressure in the DAC with sufficient accuracy to study this region, and perhaps also because the scientific importance of this region has not been widely recognized.

We have utilized both spectroscopy and diffraction to probe the EOS, structural and dynamic nature of supercritical fluid $CH_4$ from ambient pressure up to the melting curve, from 298 K to 397 K. We present a unified set of data covering both the low pressure regime (in which the repulsive inter-atomic potential competes with attractive van der Waals forces) and the high pressure regime (in which the repulsive inter-atomic potential dominates). The Widom lines are not expected to exist beyond $T / T_c \approx 2.5$ and $P / P_c \approx 3.0$[6]. As $CH_4$ has $T_c$ of 190 K and $P_c$ of 4.6 MPa[30], our investigations probe pressures beyond where the Widom lines for $CH_4$ are expected to lie and at pressures which are applicable to the interiors of the gas giants Uranus and Neptune which are rich in $CH_4$.

## II. EXPERIMENTAL AND ANALYSIS METHODS

Donut (for experiments at ambient temperature) and piston-cylinder (for experiments at high temperature) diamond anvil cells (DACs) were constructed in-house. The DACs were equipped with diamonds having 600 μm and 450 μm diameter culets, and stainless steel gaskets. $CH_4$ was liquefied inside a cryogenic loading apparatus by cooling the apparatus with liquid $N_2$. The DACs were then closed whilst completely immersed in liquid $CH_4$. The liquid $N_2$ does not enter the cryogenic loading apparatus during this procedure, and the strong Raman-active $N_2$ vibration was not observed at any point during our experiments.

To perform the high temperature experiments, the DACs were heated using a resistive heater surrounding the entire DAC (Watlow inc.) and temperature was controlled with a precision of ± 2 K using a custom-constructed temperature controller. Temperature was measured using K type thermocouples in contact with the diamond or gasket.

Pressure was measured at ambient temperature using the ruby photoluminescence method, resulting in a typical error of ± 0.002 GPa. In all cases error bars are too small to display. At temperature above 300 K pressure was measured using the Sm:YAG photoluminescence method using the Y1 peak[31], and was measured before and after the collection of each data point due to the greater risk of pressure fluctuation at high temperature. The typical error in each individual pressure measurement is ± 0.003 GPa, however some pressure fluctuation occurred whilst the $CH_4$ Raman spectra were being collected. The error bars displayed on figure 3 are the result of this fluctuation. Except where otherwise stated, data was collected on pressure decrease at constant temperature.

X-ray diffraction patterns from fluid $CH_4$ were collected on beamline ID27 at the European Synchrotron Radiation Facility. The beam size was 3 µm x 2 µm and Soller slits were utilized to reduce the background signal coming from the diamond anvils. The background (a diffraction pattern collected from the empty DAC) was subtracted from each integrated pattern after normalization in the high-$q$ limit (figure 1).

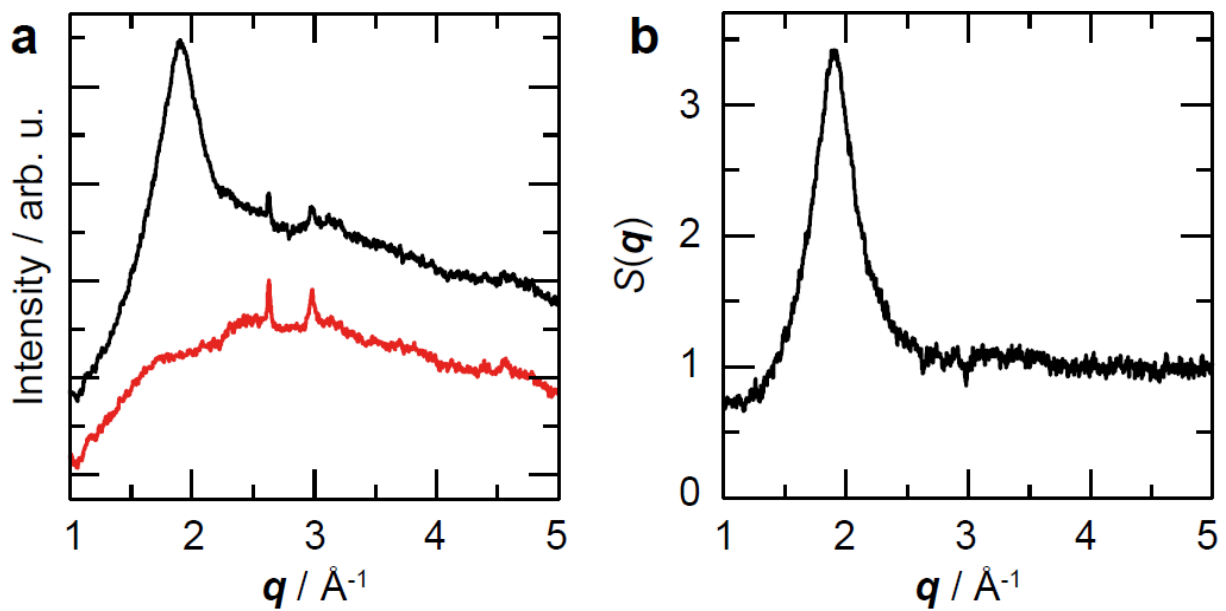

Figure 1 (color online). Background correction of integrated diffraction patterns. (a) Diffraction pattern of fluid $CH_4$ at 0.69 GPa (black) and background from empty DAC (red). (b) Static structure factor $S(q)$ of CH$_4$ at 0.69 GPa obtained by subtracting background and normalising in the high $q$ limit.

Raman spectroscopy was performed using 532 nm laser excitation, backscattering geometry and a conventional single grating (1200 lines per inch) Raman spectrometer. The spectrometer was calibrated using the Raman peaks from silicon and diamond collected at ambient conditions. After background subtraction, the strong Raman peak from the υ$_1$ symmetric stretching vibration of CH$_4$ at ca. 3000 cm$^{-1}$ was fitted with a single Lorentzian curve, allowing the determination of the peak positions with a typical accuracy of ± 0.05 cm$^{-1}$. In all cases the error bars are too small to display. Our investigation was carried out at pressures where we would not expect to observe any rotational substructure to the peak.

All Raman spectra were collected using 180° backscattering geometry through the cylinder diamond. To ensure the greatest accuracy possible in our measurements of the intensity of the υ$_1$ peak we focussed on the same point on the diamond culet before collecting each Raman spectrum. The second order Raman band from the diamond (between 2500 cm$^{-1}$ and 2700 cm$^{-1}$) was collected in the same Raman spectrum as the υ$_1$ methane peak and the drastic changes in intensity of the υ$_1$ discussed later were clearly observed relative to the intensity of the second order diamond band intensity remaining constant.

## III. RESULTS AND DISCUSSION

We collected X-ray diffraction patterns of fluid CH$_4$ at 298 K. We observed the first peak in the fluid CH$_4$ structure factor $S(q)$ with reasonable intensity (figure 1), used its position to derive the volume occupied by the CH$_4$ molecule and plotted the EOS of fluid CH$_4$ from very close to ambient pressure, up to the melting point at 1.35 GPa[32] (figure 2a). The data is fitted with a Murnaghan EOS

(equation (1)), and with an ideal gas EOS with van der Waals correction[14] (equation (4), where $A$, $B$ and $C$ are constants).

$$V(P) = \frac{A}{P-C} - B \qquad (4)$$

Both equations provide a good fit to the data. However, it is important to note that this data does not provide an accurate measure of the density. This is because, to obtain the density from the volume occupied by a single molecule, one needs to assume a certain co-ordination number, packing arrangement and concentration of holes in the fluid. These factors are unknown.

We also plot (figure 2b) the width (half width half maximum, HWHM) of the first peak in $S(q)$. This increases significantly at extremely low pressures, consistent with the sample being in a gas-like state instead of a liquid-like state[9][14]. However, in contrast to a recent report on supercritical Ar[33], we observe no evidence for a discontinuity in the position (figure S1) or width (figure 2b) of $S(q)$ as a function of pressure. Such a discontinuity is not expected, since it is determined by the density which varies smoothly this far beyond the critical point[14][17][30].

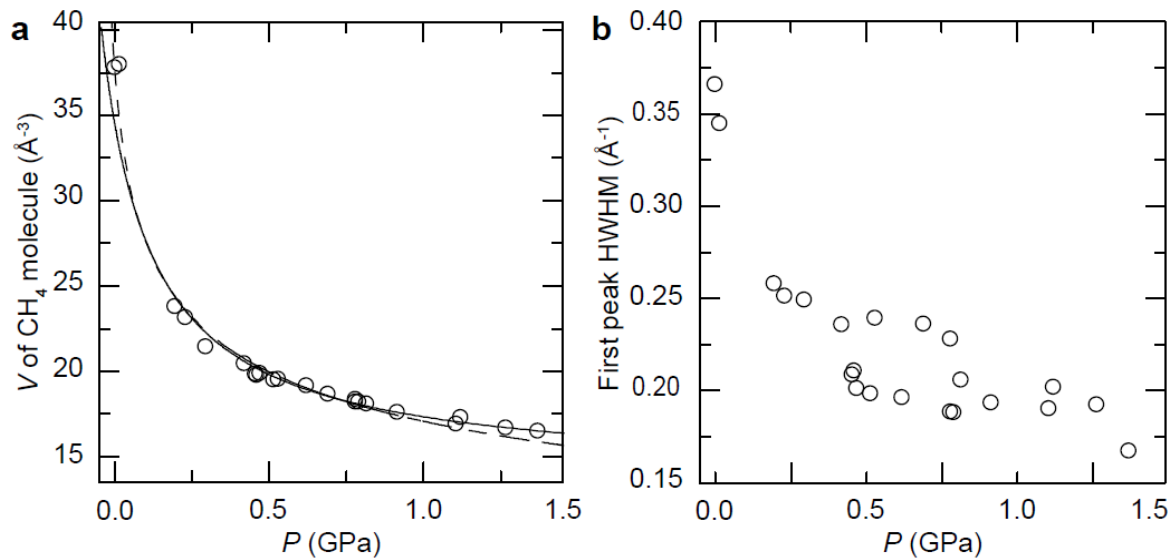

Figure 2. (a) Equation of state (EOS) of fluid CH$_4$ at 298 K. Circles: experimental data. Solid line: Fit of ideal gas EOS modified with van der Waals correction (equation (4)). Dotted line: Fit of

Murnaghan EOS (equation (1)). (b) HWHM of first peak in $S(\boldsymbol{q})$ plotted as a function of pressure at 298 K.

We attribute the large scatter in X-ray diffraction datapoints (especially the width) to relatively weak scattering from the sample and errors introduced during the background subtraction procedure.

However, using Raman spectroscopy we observe much clearer evidence for a crossover. Figure 3 shows the reduced frequency of the Raman-active $\upsilon_1$ vibration of $CH_4$ upon pressure increase at 298 K, 345 K, 374 K and 397 K. In all cases, at higher pressure the shift in vibration frequency upon pressure increase is positive and linear, as expected from a dense liquid. At lower pressure, the vibration frequency decreases slightly upon pressure increase before reaching a minimum and increasing as the higher pressure region is reached. The observed trend in peak position is reversible with no hysteresis within experimental error (see supplementary material).

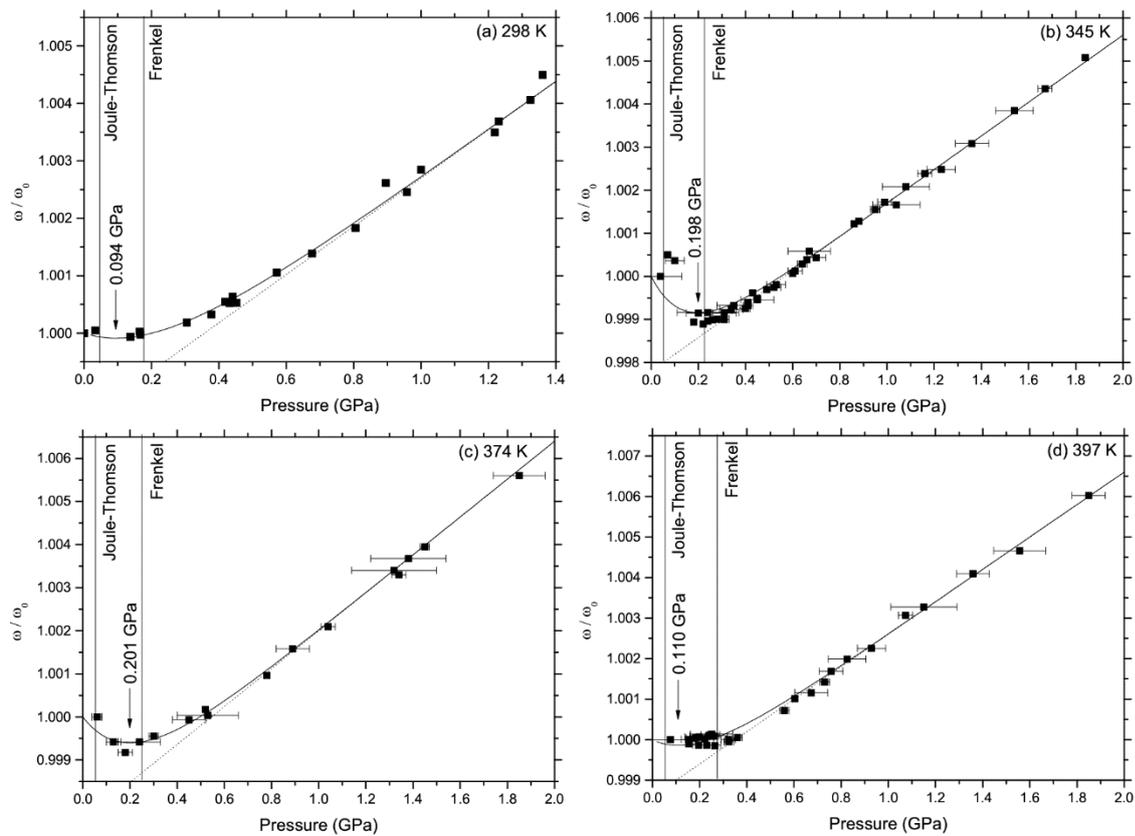

Figure 3. Plot of reduced frequency of $CH_4$ Raman-active vibration as a function of pressure at (a) 298 K (b) 345 K (c) 374 K and (d) 397 K. Solid lines following the data are fits performed using equation (6) and arrows signify the minimum of this fit. Dotted lines are linear fits to data collected above 0.6 GPa (equation (5)). Vertical solid lines are expected positions of the Joule-Thomson line and Frenkel line.

Therefore, we propose that in the high pressure region the vibration frequency is described by equation (2), *i.e.* its pressure-induced shift is entirely due to the repulsive interaction between $CH_4$ molecules, and make linear fits to the data above 0.6 GPa (using equation (5), where $c_1$ and $c_2$ are constants). This provides a good fit to the data above 0.6 GPa at all temperatures studied (298 K, 345 K, 374 K and 397 K) and $c_2$ remains constant (within error) at 0.004 GPa$^{-1}$. The value of $\omega_0$ was fixed at the observed vibration frequency for the lowest pressure datapoint at each temperature (in all cases lower than 0.1 GPa).

$$\frac{\omega(P)}{\omega_0} = c_1 + c_2 P \qquad (5)$$

However, in the low pressure region the repulsive potential (responsible for the increase in Raman frequency, equations (3) and (5)) competes with the attractive van der Waals interaction between $CH_4$ molecules, which is causing a decrease in Raman frequency as pressure is increased. Since the attractive van der Waals contribution becomes weaker as pressure is increased we describe it using an exponential decay function. The total Raman shift as a function of pressure is therefore given as follows:

$$\frac{\omega(P)}{\omega_0} = c_1 + c_2 P + (1 - c_1)e^{-c_3 P} \qquad (6)$$

Here, $c_1$ and $c_2$ are the constants which remain fixed following our fits to the data above 0.6 GPa whilst $c_3$ is adjusted so that equation (6) provides the best fit to the entire dataset. To evaluate when liquid-like behaviour becomes dominant over gas-like behaviour, we obtain the pressure at the minimum in the fit to the graph of $\frac{\omega(P)}{\omega_0}$. It is clear from figure 3 that due to the spread of data and lack of Raman datapoints at extremely low pressure this curve fit is not always well constrained.

For comparison we therefore examined our data using two other methods. We fitted straight lines to the datapoints lying on each side of the perceived "kink" in the graph of reduced Raman frequency (figure 3) and calculated the crossover pressures as where these lines intersect, as performed in the previous diffraction studies refs. [33] and [40]. This shifts our crossover pressures higher, especially at 397 K (see supplementary material). This procedure, in contrast to the fit using equation (6), is purely phenomenological. It assumes a completely discontinuous transition (none of the proposed transitions / crossovers in the supercritical fluid regime are expected to be completely discontinuous) and it involves some preconception of where the "kink" and therefore crossover pressure lies. In the fit using equation (6) there is no assumption as to where the crossover lies, or even that it exists at all - there is nothing in the curve fitting procedure to prevent $c_3$ converging to a small value or zero. This did not happen.

We have also plotted on figure 3 the theoretically expected positions of the Joule-Thomson curve and Frenkel line in $CH_4$, for direct comparison to our actual Raman datapoints (see later discussion).

Since the procedure of using only linear fits is purely phenomenological and the direct comparison of our datapoints to theoretically predicted crossovers does not produce a numerical value for the experimentally observed crossover we use equation (6) for our determinations of the crossover pressure. Whilst the curve fits are poorly constrained in this study, the procedure can provide a framework for future investigations. These fits at all temperatures studied are shown in figure 3.

We observe an additional phenomenon at pressure close to the minimum in each case; a discontinuous change in the integrated intensity and width of the Raman peak. Figure 4a shows this at 298 K. This change in intensity, peak position and width is also reversible at all temperatures studied (see supplementary material). It remains unclear exactly how the transition from the liquid-like to gas-like behaviour causes this phenomenon; however we can see possible links. For instance, the propagation of shear waves on the liquid-like side of the Frenkel line could cause localized compression of some areas of the fluid with a resulting broadening of the Raman peak.

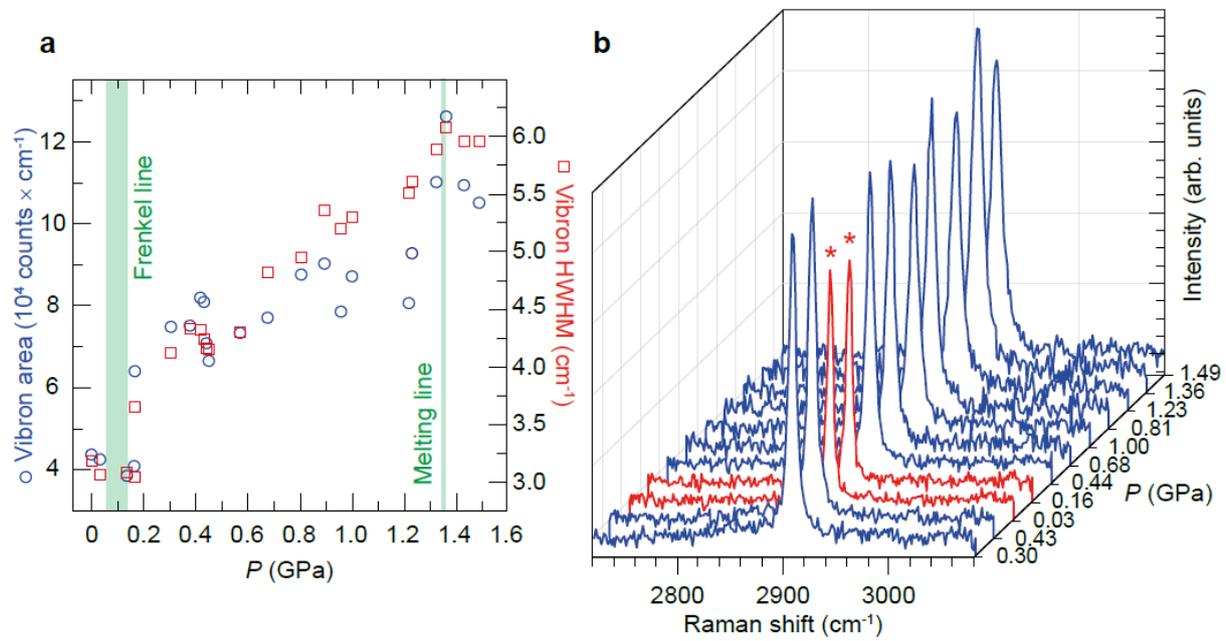

Figure 4. (color online) (a) Plot of Raman peak area (integrated intensity, blue circles) and HWHM (red squares) as a function of pressure at 298 K, showing discontinuous decrease in both area and HWHM when the gas-like side of the Frenkel line is reached. The green shaded region is the area in which the minimum in vibration frequency (fitted using equation (6)) lies, which we associate with the Frenkel line and the green line marks the experimentally measured melting line[32]. (b) Waterfall plot of spectra collected at 298 K in order collected, demonstrating the reversible nature of the crossover. Red spectra (marked with asterisks) are those on the gas-like side of the Frenkel line.

In figure 5 we plot the crossover pressures found using the procedure above (equation (6)), at all temperatures studied. The crossover shifts to higher pressure as temperature is increased as expected, but we note the significant scatter in the data due to the poorly constrained curve fits using equation (6). Whilst the crossover pressure obtained at 397 K using equation (6) is at somewhat lower pressure than that at 374 K, the alternate methodology of fitting straight lines to the data above and below the perceived "kink" in each dataset results in a crossover pressure of 0.36 GPa at 397 K (see supplementary material).

The given errors in crossover pressures in figure 5 are calculated from the errors on the fitted parameters $c_2$ and $c_3$ in equation (6). However, we expect the real error in these pressures is slightly larger. This is due to the method of pressure measurement; photoluminescence from a Sm:YAG crystal in the sample chamber. At high temperature, the photoluminescence peak used to measure pressure broadens and merges with the neighbouring peak (see example spectra in supplementary material figure S7). This produces an error in pressure measurement which is hard to quantify, and is what limited the temperature achievable in this study. We recommend the development of pressure sensors providing more accurate pressure measurement above 400 K at modest pressures, to address this problem.

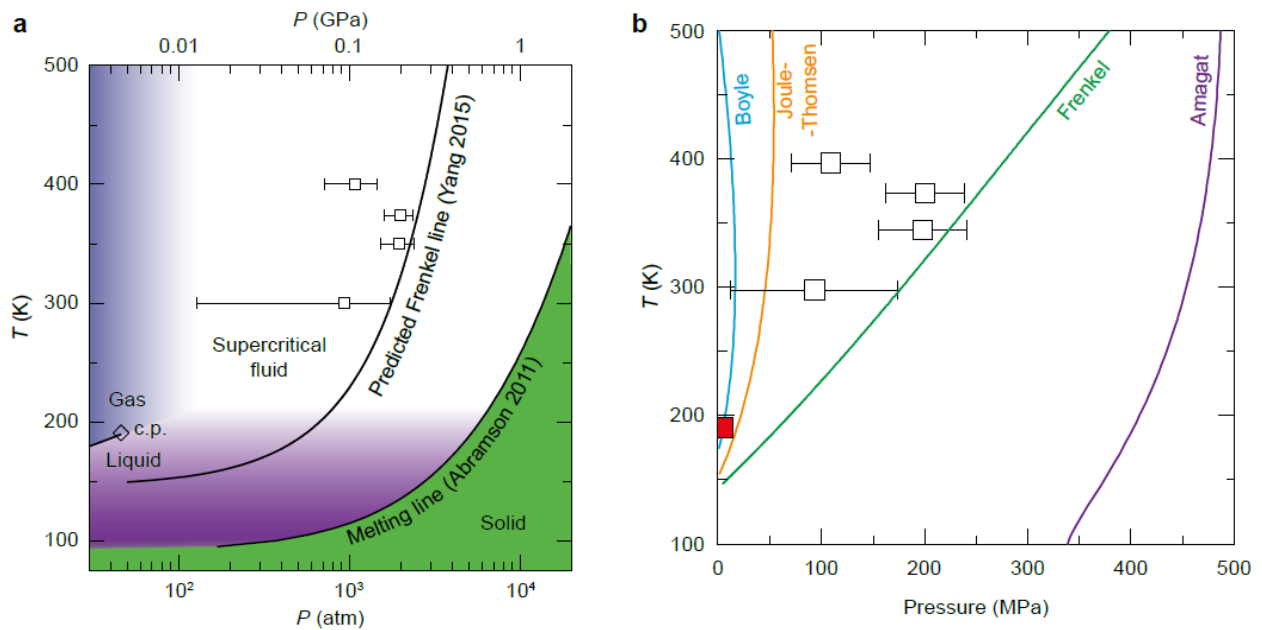

Figure 5 (color online). Phase diagram of $CH_4$. (a) The theoretically predicted Frenkel line in $CH_4$[16] is marked along with the experimentally observed melting curve[32]. The data points are the crossovers we observe experimentally and attribute to the Frenkel line. (b) The position of the theoretically predicted Frenkel line compared to the Boyle curve, Joule-Thomson curve and Amagat curve. Open squares are our Frenkel line datapoints and the red closed square is the critical point (c.p.).

Despite limitations imposed by the need for accurate pressure measurement, we have succeeded in observing a crossover in $CH_4$ at $44P_c$ using equation (6) ($77P_c$ using the alternate fitting methodology outlined in the supplementary material) and $2.1T_c$. The Widom lines are not expected to persist up to this pressure and we have confirmed this for the Widom lines in $CH_4$ (see supplementary material and ref. [30]).

We therefore compare our crossover pressures to the expected positions of the Boyle curve, Joule-Thomson curve and Amagat curve (calculated using the Wagner-Setzmann equation of state for methane and the ThermoC package [34]) and the expected position of the Frenkel line[16] (figure 5b). Out of these options, only the Joule-Thomson curve and Frenkel line lie close enough to our observed crossover to be feasible candidates to explain it.

In terms of pressure, our crossover pressures (when obtained using equation (6) as shown in figure 5, or using linear fits as shown in the supplementary material) lie, on average, closer to the Frenkel line than the Joule-Thomson curve. Alternatively, we could discuss the data in terms of density. Since, at higher pressure, a particular increase in pressure produces a smaller increase in density, this would shift our crossover pressures even closer to the Frenkel line than the Joule-Thomson curve. In terms of temperature, our crossovers lie at pressures where the Joule-Thomson curve is not predicted to exist, at any temperature.

A direct comparison of the expected positions of the Frenkel line and Joule-Thomson curve to our plots of Raman frequency against pressure (figure 3) reveals that the expected Frenkel line lies, at all temperatures measured, closer to the point where the Raman frequency stops decreasing upon pressure increase and starts increasing, than the Joule-Thomson curve.

The predicted Amagat curve lies at significantly higher pressure than the Frenkel line (over 3x the pressure of the Frenkel line at 300 K). The prediction of the Amagat curve, just like that of the Boyle curve and Joule-Thomson curve, stems from an understanding of the supercritical fluid state as a dense non-ideal gas. We argue, therefore, that the predicted Amagat curve is not meaningful on account of the fact that it lies at significantly higher pressure than the Frenkel line. In this P-T region we propose that the fluid must be treated as a solid in which the molecules are relatively closely packed with specific positions and only occasional vacancies.

We propose that the Frenkel line is the most plausible explanation for our observations, on account of the fact that it lies the closest to our datapoints, and because the behaviour of the Raman-active vibration and first $S(q)$ peak on the low pressure side of the crossover we observe are as expected for a gas-like sample while their behaviour on the high pressure side of the crossover are as expected for a liquid-like sample. This is the distinction that the Frenkel line makes. Furthermore, we observe (see figure 4) a discontinuous, reproducible and reversible change in the intensity and width of the Raman peak at pressure close to the minimum in each case. There are plausible mechanisms by which this can be associated with the Frenkel line (for instance, the propagation of shear waves through the fluid on the liquid-like side could cause local variation in density and hence broadening), but we can see no mechanism by which the observed discontinuous change could be associated with the Joule-Thomson line.

## IV. CONCLUSIONS

We have observed a narrow crossover between liquid-like and gas-like behaviour in $CH_4$ at 400 K; 44 – 77 $P_c$ (depending on the curve fitting method) and 2.1$T_c$. We propose the Frenkel line as the most likely explanation for this observation. The observation of this crossover in a simple molecular fluid has major implications for planetary science. In particular concerning Uranus, where the atmosphere is thought to be isolated from its interior, evidenced by anomalously low heat flow measurements from the Voyager spacecraft[35] and storm activity attributed to seasonal change in incident solar flux[36]. These findings have been attributed to the remnant heat of Uranus being trapped beneath a less conductive layer[37]. We believe[11][6][38] that the thermal conductivity and other heat-related properties of fluids change significantly when the Frenkel line is crossed, and that this is likely to occur in the transition zone between the atmosphere and mantle of Uranus, and Neptune. Furthermore, understanding the Frenkel line in other dominant planetary species, (hydrogen, helium, water and ammonia), would improve our understanding of all gas giants. We believe the findings presented here will be relevant to characterization of supercritical fluid inclusions in rocks[24][39] and industrial use of supercritical fluids.

This work also calls for the liquid state to be re-examined. A region in P-T space exists (figure 5a) where a sample exists on the liquid-like side of the boiling curve (so it is a liquid in terms of static properties) but on the gas-like side of the Frenkel line (so it is a gas in terms of dynamic properties). This region warrants careful study using modern experimental techniques.

We must acknowledge the growing body of evidence demanding a change in our understanding of the supercritical fluid state, despite the issue attracting controversy[17][33]. In addition to our observation of the Frenkel line here, a crossover has been observed using molecular dynamics simulations in Ar at $T / T_c$ = 3.0, $P / P_c$ = 100 (attributed to a Widom line[4], and alternatively to the Frenkel line[6]). A crossover was observed very recently in Ne at $P / P_c$ = 250 and $T / T_c$ = 6.6[40] and attributed to the Frenkel line. These works have observed crossovers in static[40] and dynamic[4] structure factors corresponding to the Frenkel line, whilst here we observe a crossover using Raman spectroscopy. The Raman frequency, as discussed above, depends on both the density and dynamics of the fluid. It exhibits a sharp crossover in all our experiments, which we associate with the Frenkel line. We conclude that the experimental evidence demonstrating that the supercritical fluid state is not a single homogeneous state is now irrefutable.

**Acknowledgements**

X-ray diffraction data was collected at beamline ID27, ESRF (beamtimes CH-4114, CH-4386). VVB is grateful to the RSF for financial support (grant number 14-22-00093). We would like to acknowledge the assistance of Dr. Volodymyr Svitlyk and Dr. Mohammed Mezouar (ESRF), Prof. E. Gregoryanz (University of Edinburgh), and useful discussions with Prof. Ian Morrison (University of Salford), Dr. Kostya Trachenko (Queen Mary University of London), Dr. Clemens Prescher (Universität zu Köln) and the anonymous referee. We would like to acknowledge the assistance of technical support staff at the University of Hull (Nigel Parkin) and University of Salford (Michael Clegg) for construction of the DACs used in this work, and Ph.D. scholarships at the University of Hull and University of Salford.

**Supplementary material**

**X-ray diffraction data**

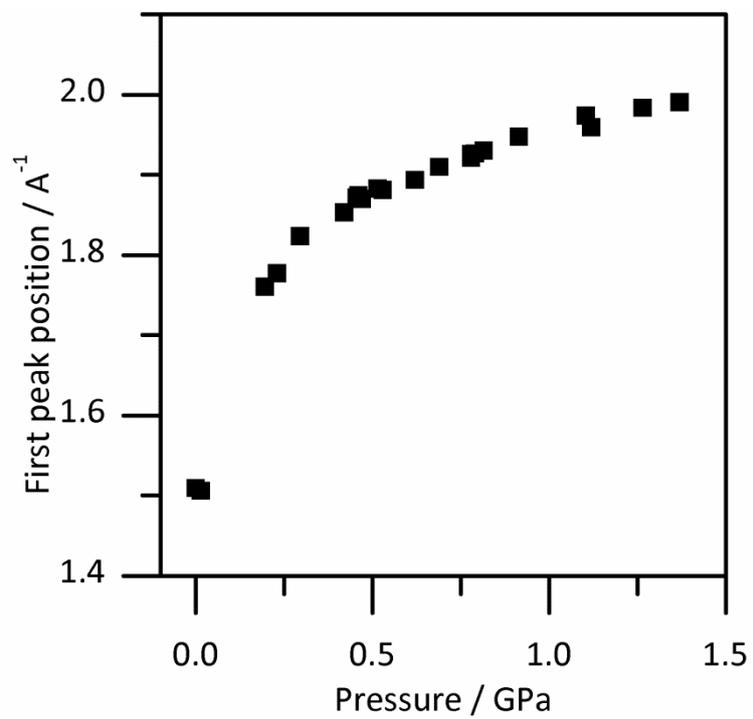

Figure S1. Plot of X-ray $S(q)$ peak position as a function of pressure at 298 K.

**Raman data**

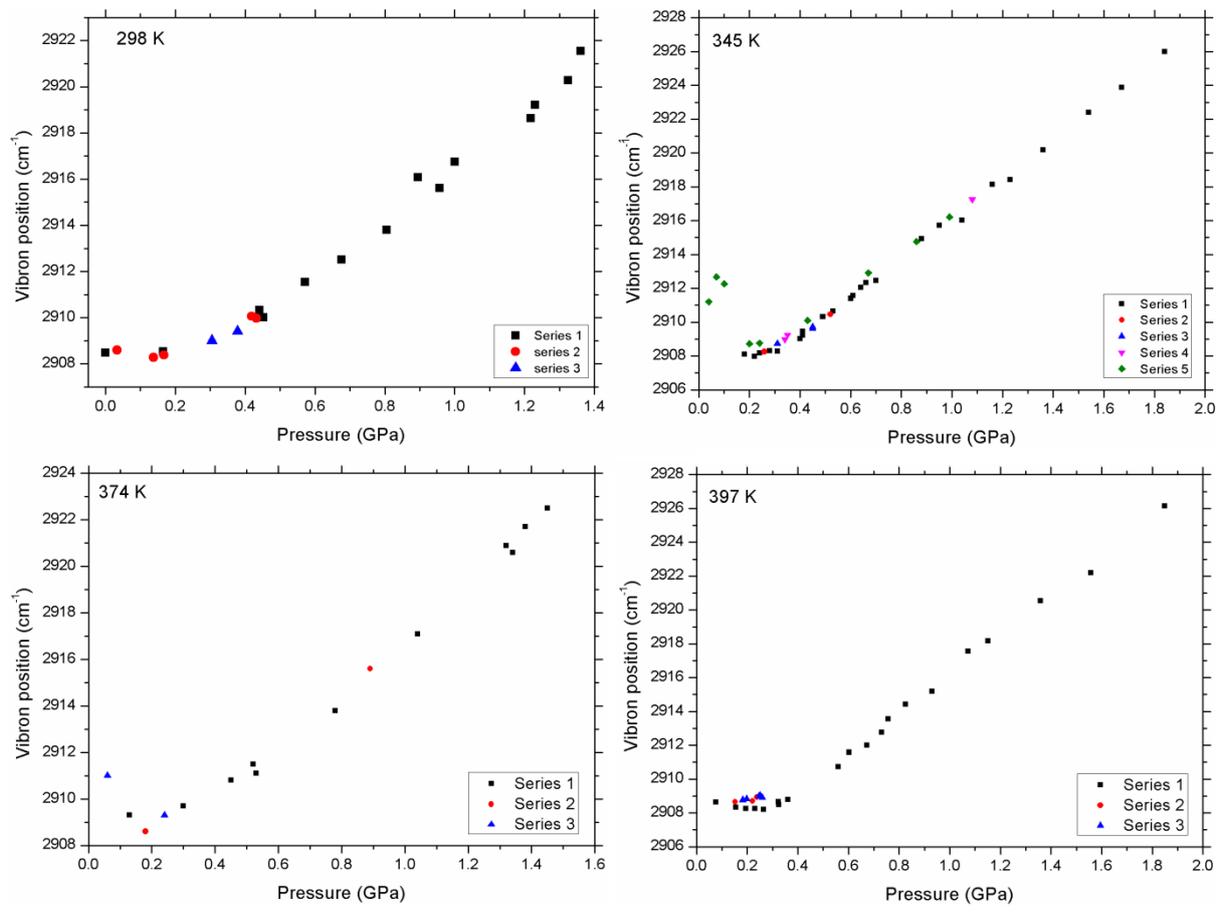

Figure S2. Plot of Raman peak position $\omega$ as a function of pressure, at all temperatures studied. At each temperature, data was collected in a single experiment upon pressure decrease (series 1), then pressure increase (series 2), then pressure decrease (series 3), to demonstrate the reversibility of the changes observed. At 345 K pressure was increased then decreased again (series 4 and 5).

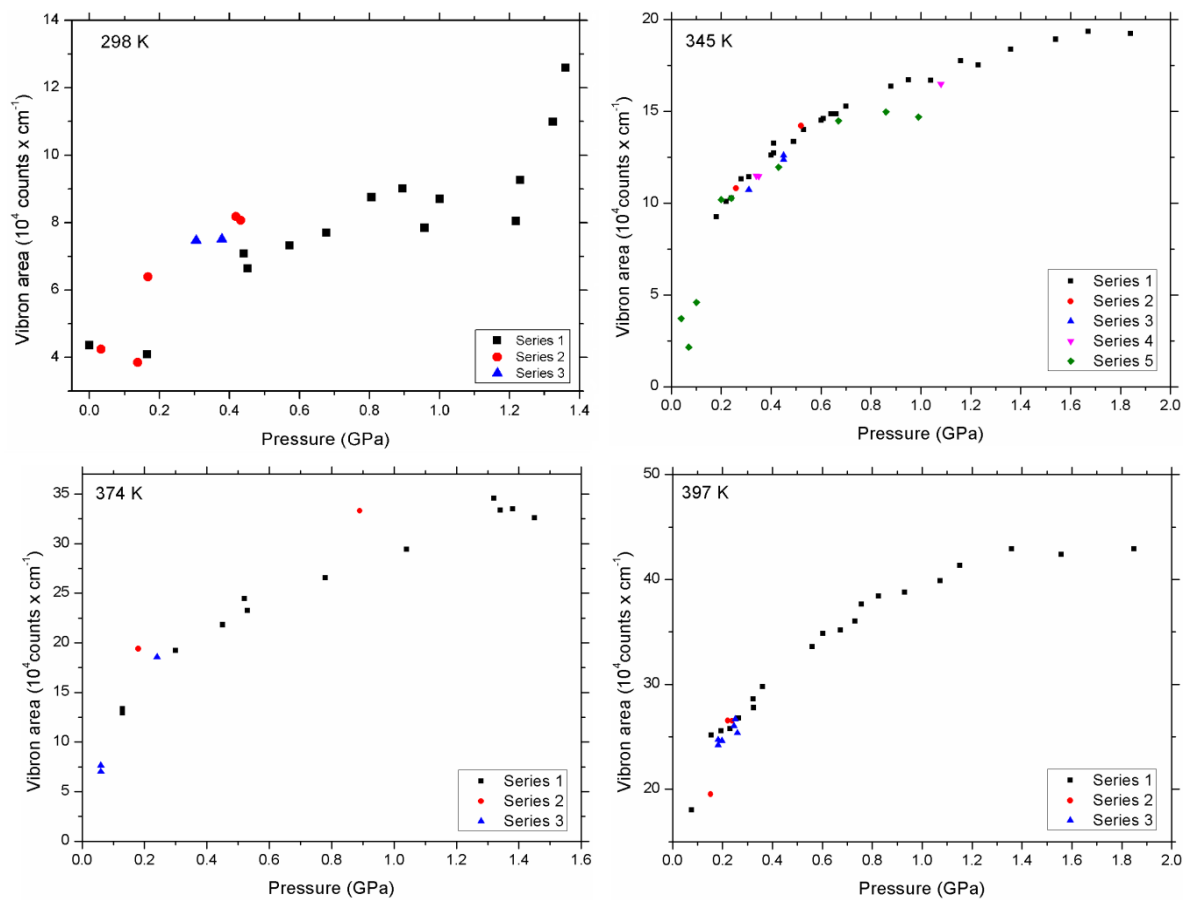

Figure S3. Plot of Raman peak area as a function of pressure, at all temperatures studied. At each temperature, data was collected in a single experiment upon pressure decrease (series 1), then pressure increase (series 2), then pressure decrease (series 3), to demonstrate the reversibility of the changes observed. At 345 K pressure was increased then decreased again (series 4 and 5).

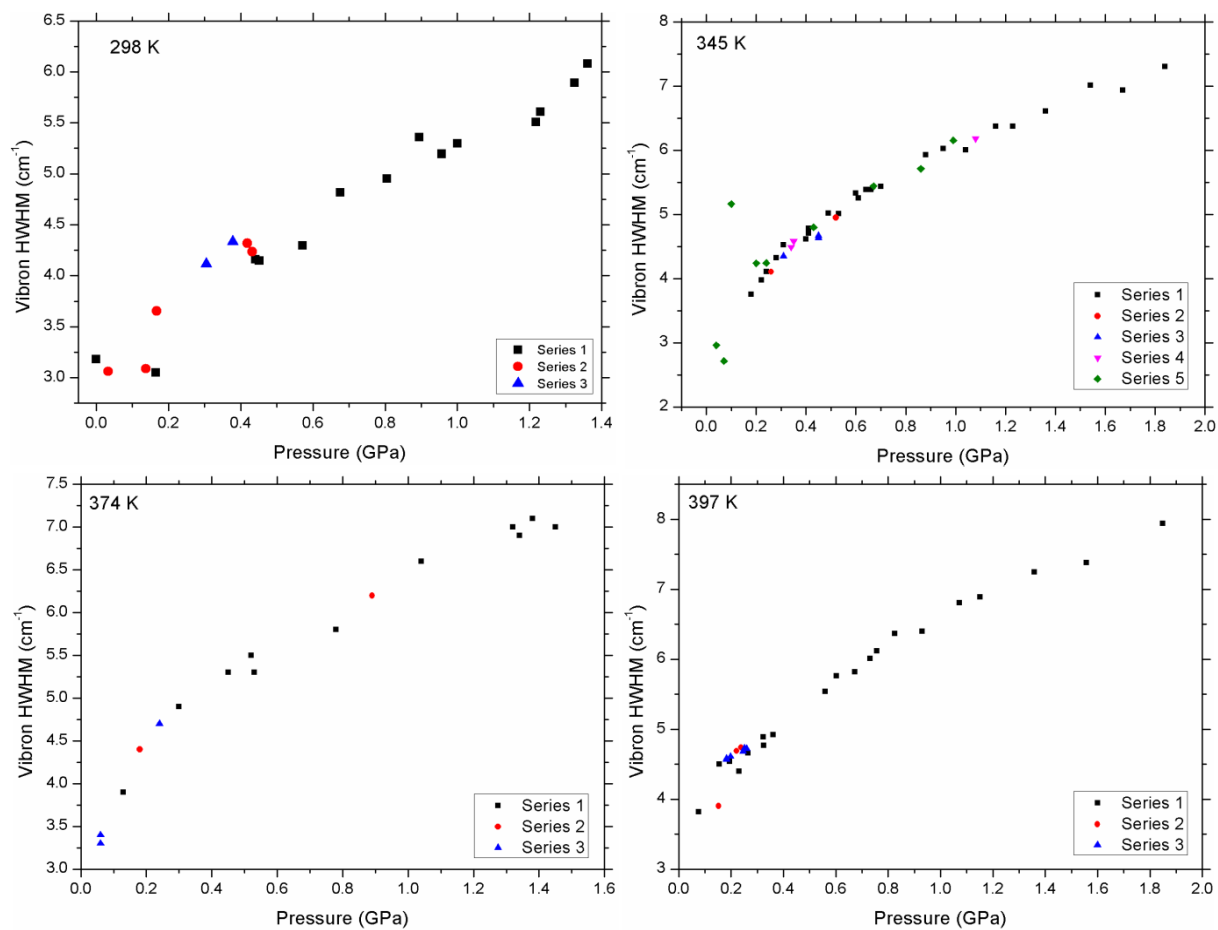

Figure S4. Plot of Raman peak half width half maximum (HWHM) as a function of pressure, at all temperatures studied. At each temperature, data was collected in a single experiment upon pressure decrease (series 1), then pressure increase (series 2), then pressure decrease (series 3), to demonstrate the reversibility of the changes observed. At 345 K pressure was increased then decreased again (series 4 and 5).

**Linear fits to graphs of reduced Raman frequency**

In addition to the fits using equation (6) to the graphs of reduced Raman frequency versus pressure we performed linear fits to the data on the high pressure and low pressure side of the observed change of gradient. These fits are shown in figure S5 (analogous to figure 3 in the main text). Figure

S6 compares the transition pressures obtained by calculating the intersects of these linear fits with those obtained using equation (6) in the main text, and the expected Frenkel line.

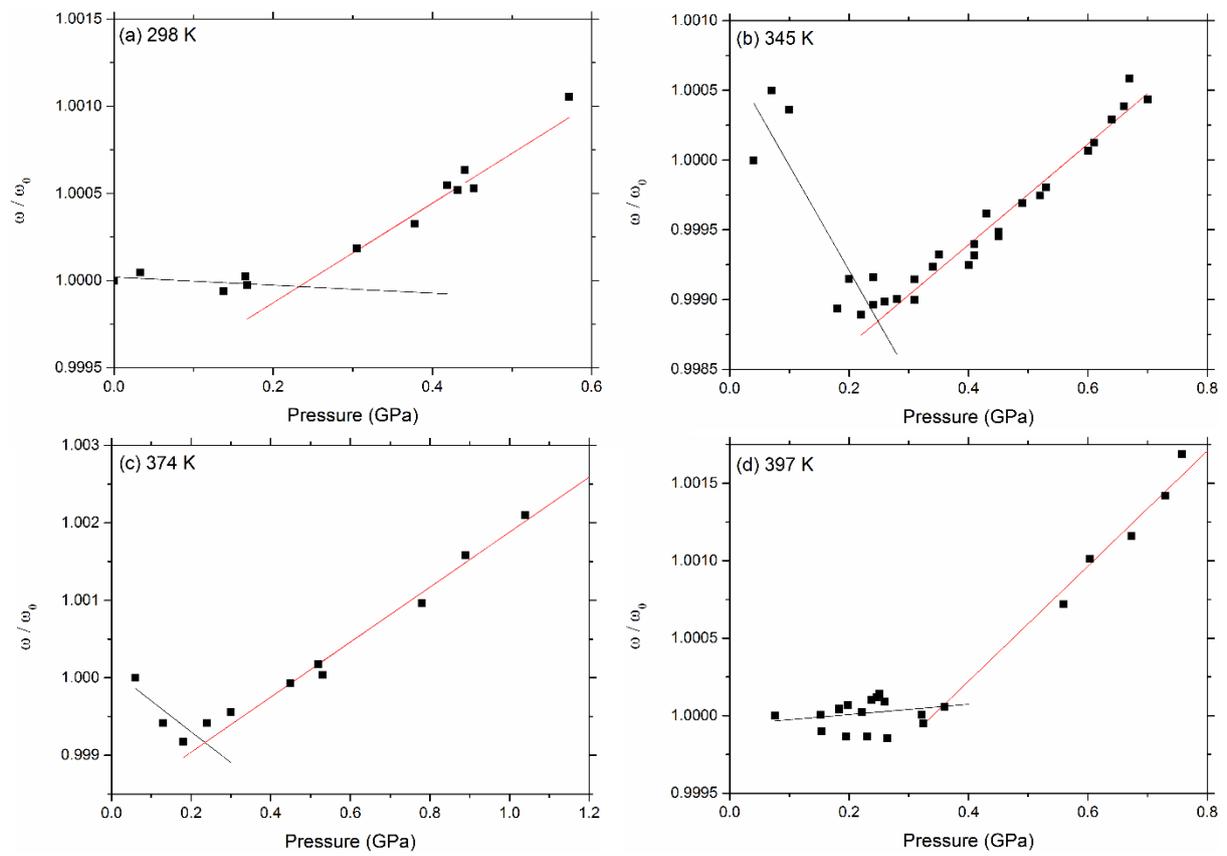

Figure S5. Plots of reduced frequency of $CH_4$ Raman-active vibron analogous to figure 3 in the main text, with linear fits to the data on the high pressure and low pressure side of the observed change of gradient.

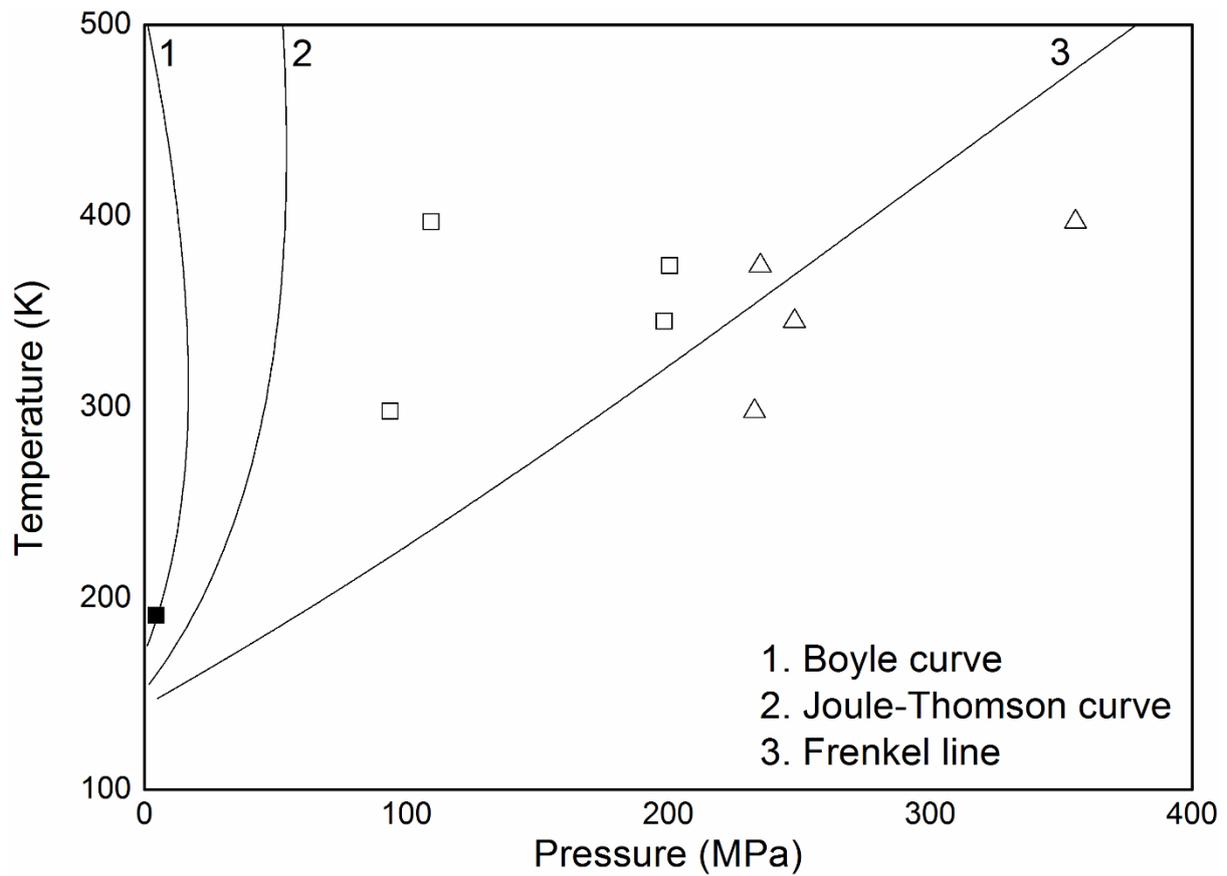

Figure S6. Theoretically predicted Frenkel line and Joule-Thomson curve compared to transition pressures obtained from our data using equation (6) in the main manuscript (open squares) and using linear fits (closed squares).

**Pressure measurement**

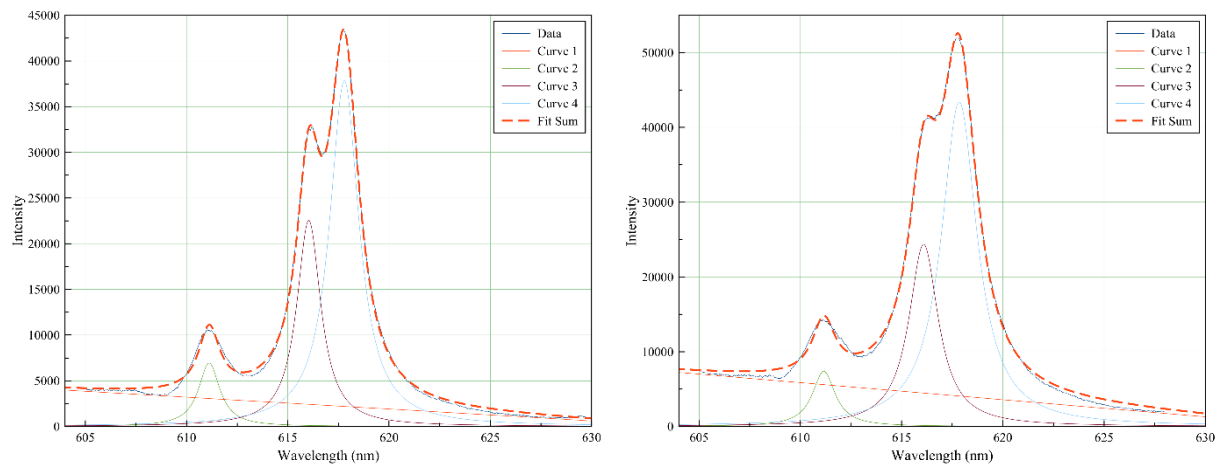

Figure S7. Photoluminescence spectra of Sm:YAG used to measure pressure, collected at 345 K, 0.52 GPa (left) and 397 K, 0.56 GPa (right). In both spectra curve 4 (the fit to the Y1 peak) is used to measure pressure.

**Widom lines in CH$_4$**

We have plotted in figures S8 the isochoric ($C_p$) and isobaric ($C_v$) heat capacities of CH$_4$ up to 400 MPa just above the critical point at 200 K, and much further beyond the critical point at 300 K. The data plotted is that available from the NIST Webbook [30]. Figure S9 shows equivalent data for the speed of sound in CH$_4$. In these cases (and other parameters available on NIST) there is a distinct extremal value at 200 K which has become smeared out by 300 K. Only $C_v$ displays any extremal value by 300 K and even that is rather smeared out. The maximum exists at 21 MPa, a far lower pressure than the Frenkel line.

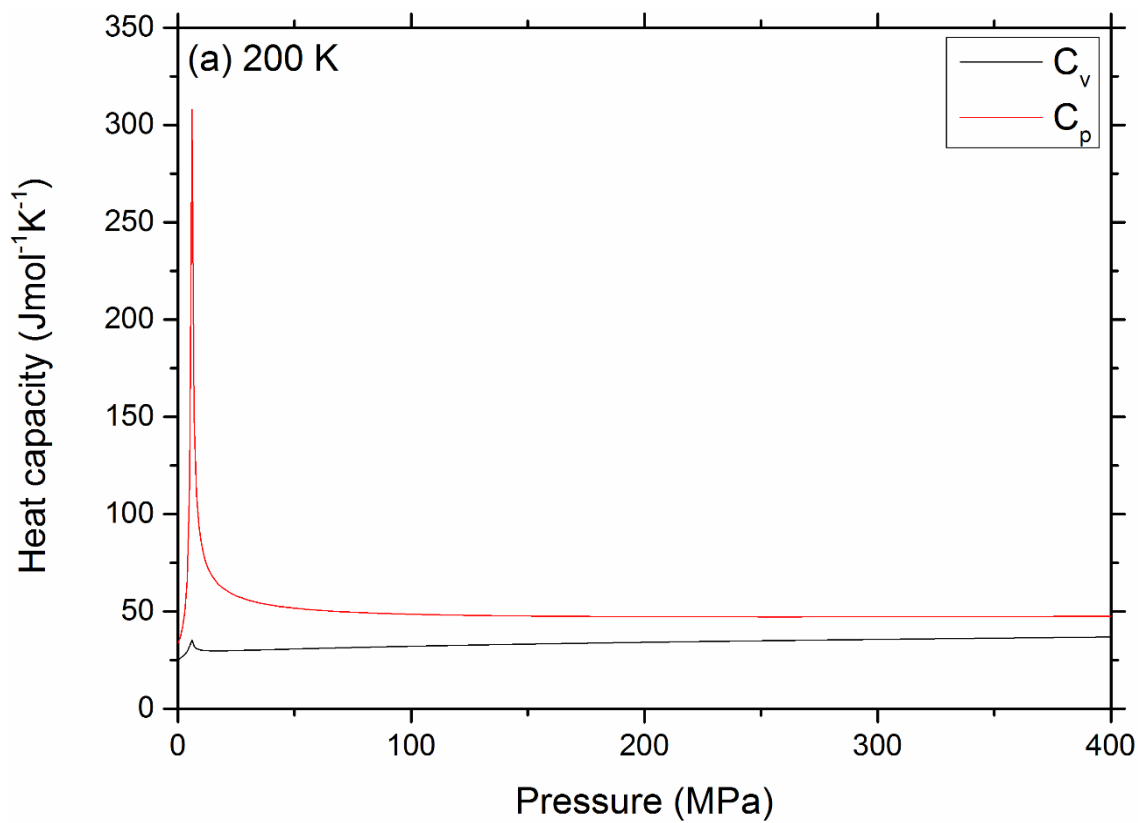

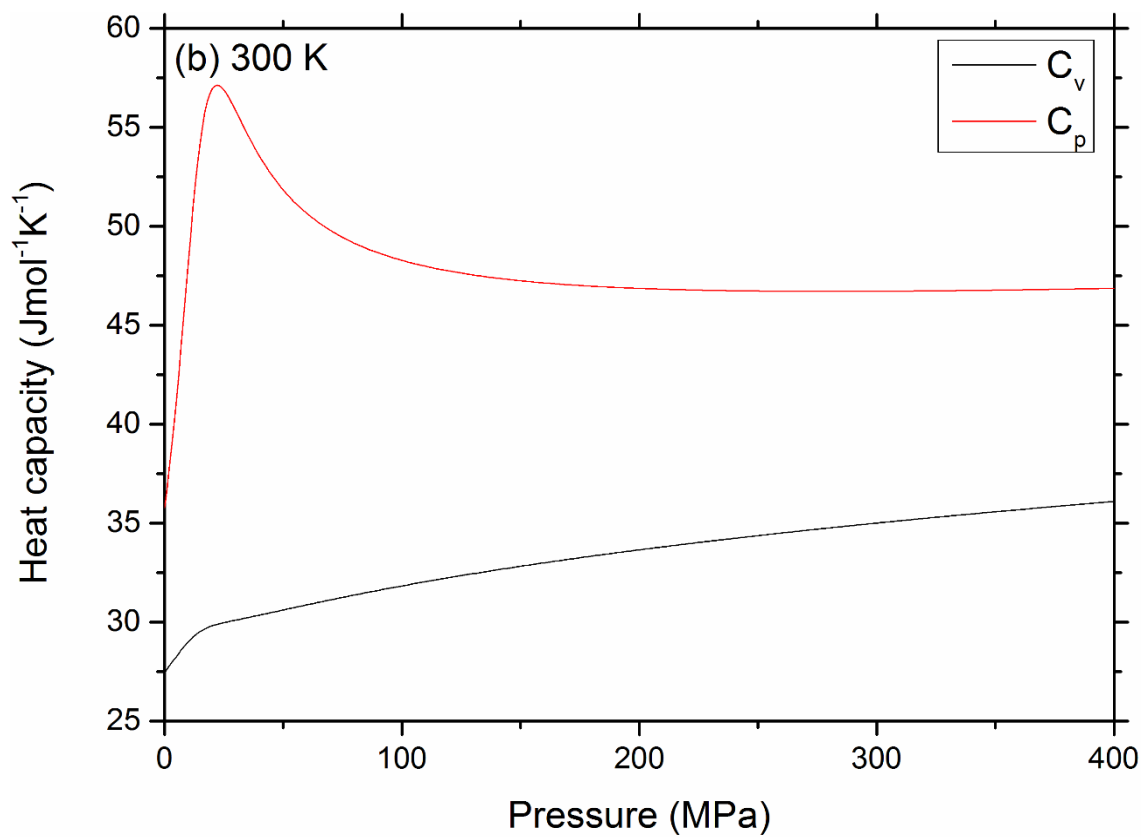

Figure S8. (a) Plots of $C_v$ and $C_p$ in $CH_4$ at 200 K. The maxima mark points on the Widom line, emanating from the critical point of 4.5 MPa, 190 K. (b) Plots of $C_v$ and $C_p$ at 300 K. Only $C_p$ displays a maximum, that is rather more smeared out than at 200 K and located at 21 MPa.

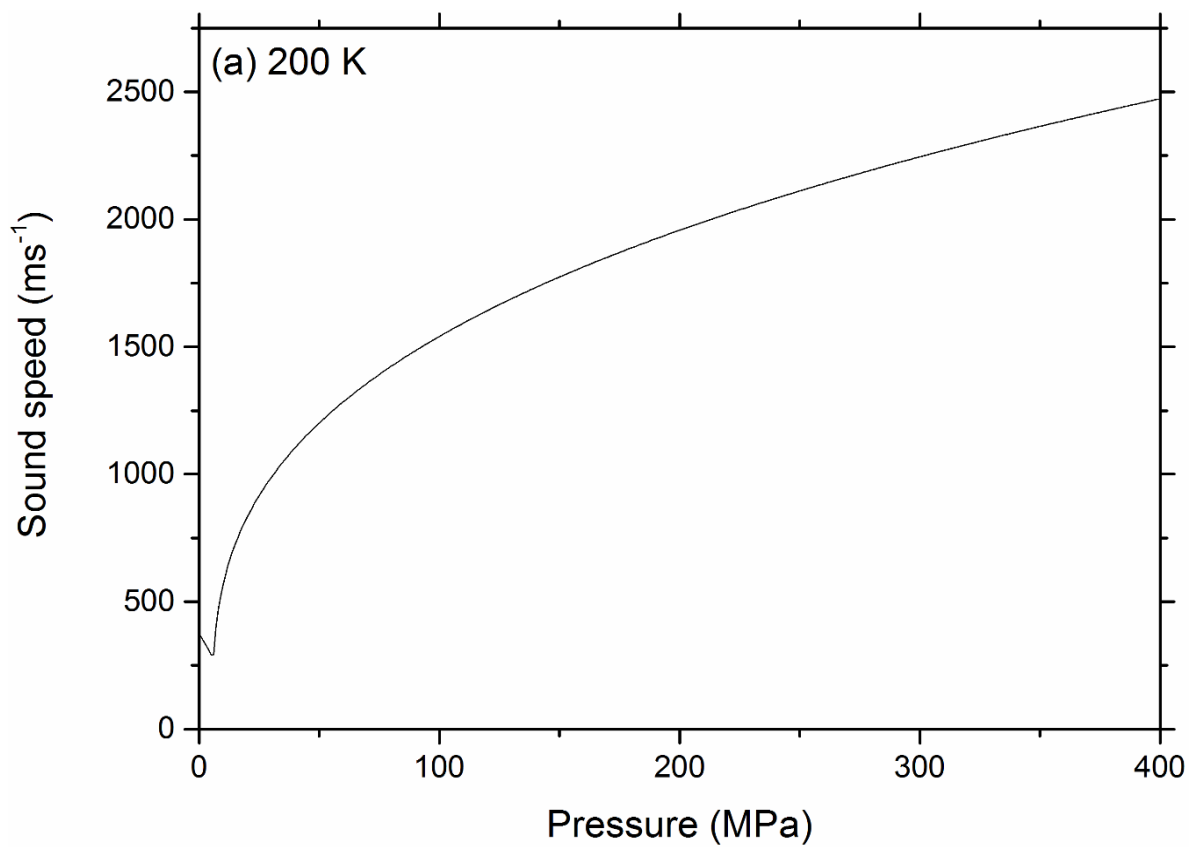

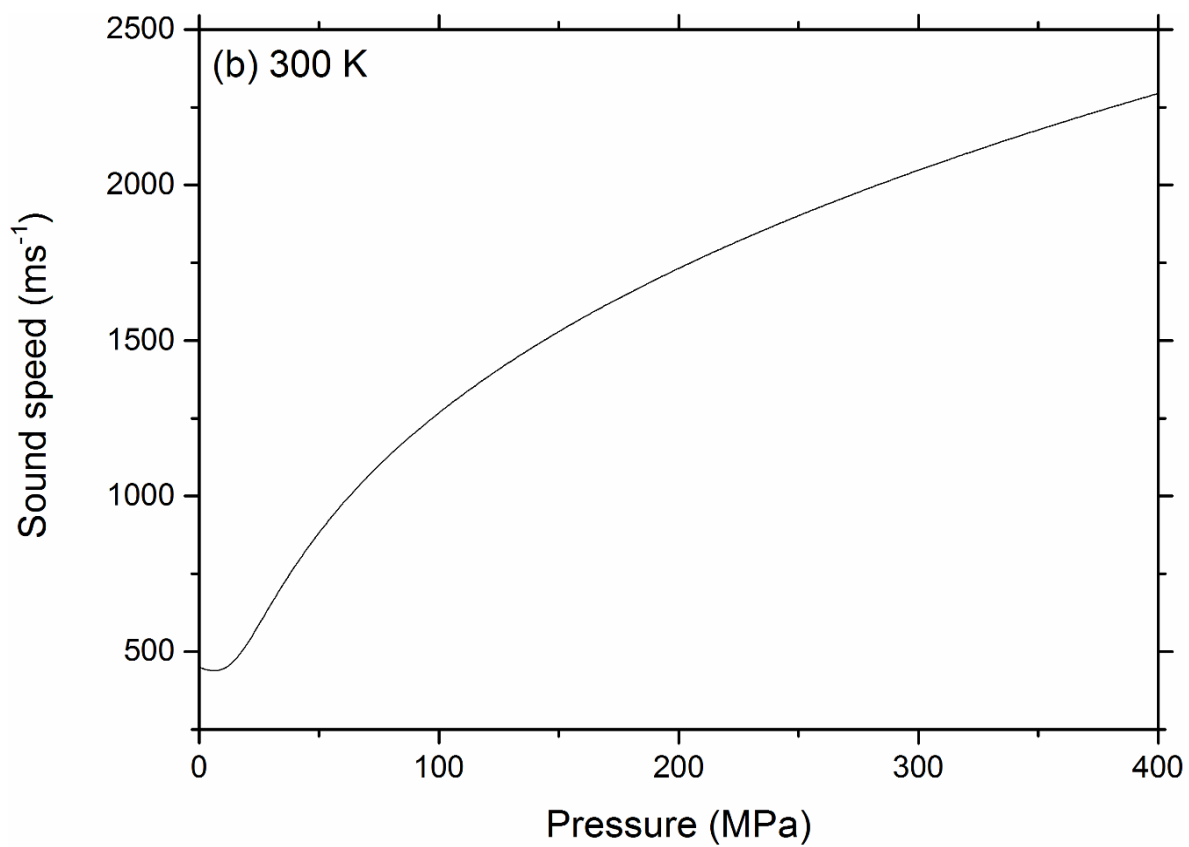

Figure S9. (a) Plot of the speed of sound in $CH_4$ at 200 K. The minimum marks a point on the Widom line, emanating from the critical point of 4.5 MPa, 190 K. (b) Plot of the speed of sound in $CH_4$ at 300 K. The minimum is almost completely smeared out by the point, and is found at 6 MPa.